\documentclass{optica-article}

\usepackage{lineno}
\usepackage{siunitx}
\usepackage{cleveref}
\usepackage{float}
\journal{opticajournal} 

\articletype{Research Article}

\begin{document}
    
	\title{Beamfit: Algorithmic Wavefront Reconstruction of Laser Beams Using Multiple Intensity Images and Laguerre- or Hermite-Gaussian Basis.}
	\author{Kevin Weber,\authormark{1,2,*} Jonathan Joseph Carter,\authormark{1,2} Sina Maria Koehlenbeck,\authormark{1,2} Gudrun Wanner,\authormark{1,2} Gerhard Heinzel\authormark{1,2}}

	\address{\authormark{1}Leibniz University Hannover, Institute of Gravitational Physics, Callinstraße 38, 30167 Hannover, Germany\\
			\authormark{2}Max Planck Institute for Gravitational Physics (Albert Einstein Institute), Callinstraße 38, 30167 Hannover, Germany\\
	}

	\email{\authormark{*}kevin.weber@aei.uni-hannover.de}

	\begin{abstract*}
		Wavefront errors are a common artifact in laser light generation and imaging. They can be described as an aberration from the spherical wavefront of an ideal Gaussian beam by combinations of higher-order Hermite- or Laguerre-Gaussian terms. Here, we present an algorithm called Beamfit to estimate the mode composition from a series of CCD images taken over the Rayleigh range of a laser beam. The algorithm uses a user-defined set of Hermite- or Laguerre-Gaussian modes as the basis of its theoretical model. A novel method reduces the number of calculations needed to compute the model's intensity profiles. For a given model containing $N$ modes, the number of Hermite-Gaussian complex amplitudes needed to calculate are reduced from orders of $\mathcal{O}(N^2)$ to $\mathcal{O}(N)$ and replaced by simple multiplications. Additionally, non-beam parameters are pre-calculated to further reduce the search space dimension and its resulting calculation time. It is planned to release the Beamfit software to the public under an open-source license.
	\end{abstract*}
	
    \section{Introduction}
    Beam quality measurements for monochromatic light sources are widely used across many disciplines in physics. Some of those methods are, for example, the M$^2$ factor \cite{Johnston1998}, power in the bucket method \cite{Chen2021}, and Strehl ratio \cite{Roberts2004}, to mention a few. They are essential diagnostic tools to estimate a laser beam quality in terms of its spatial distribution and can give insight into noise contributions stemming from imperfections in their profile. 
    Especially wavefront characterizing methods, experimentally or computationally, find use in multi-mode fibers \cite{Li2017, Choi2019}, high-power laser characterization \cite{Wellmann2019, Chen2021}, and gravitational wave detectors \cite{Schiworski2021}. In space-based interferometers, deformations in laser beams cause phase front errors that can couple directly into the readout via, e.g., pointing jitter as a major noise source \cite{Sasso2018, Xiao2023}. Therefore, carefully characterizing such beams is crucial to ensure the success of these missions. More information than is contained than the widely used M$^2$ factor would be required in these cases.
 
    For inter-satellite interferometers, beams are usually propagated over hundreds \cite{gfo2012, Sheard2012}, if not millions \cite{Danzmann2017, Xiao2023} of kilometers. Recreating these conditions in a laboratory environment is unfeasible. Therefore, accurate laser beam simulations are crucial to predict such an instrument's performance. The beam's physical properties must be described as realistically as possible to extrapolate them faithfully over large distances. An approach to satisfy these prerequisites is the modal decomposition method \cite{Kwee2007, Schulze2012a, Schulze2013, Kotlyar1998, Ma2009, Cox2019, Huang2015, Choi2019, Schiworski2021}. It is a versatile tool that has been already well-studied. Laser beam simulations can use decomposition methods to efficiently simulate the propagation of complicated wavefronts through apertures and optics.
    
    For experimental decomposition, a highly specialized setup is usually needed to perform the decomposition. Such beam-decomposing setups can consist of optical ring resonators \cite{Kwee2007}, computer-generated holograms \cite{Schulze2012a}, spatial light modulators \cite{Schulze2013}, phase-diffractive optical elements \cite{Kotlyar1998}, digital micro-mirror devices \cite{Cox2019}, or low-contrast interferometer \cite{Ma2009}. These experimental methods are highly efficient at decomposing to a mode basis and can usually decompose modes up to very high orders in real-time. However, this requires costly laboratory equipment, which usually takes time to set up and calibrate.
    
    For the computational approach, only an intensity- or phase-profiling camera to record the beam's spatial distribution is needed. That gives these methods high versatility at the cost of lower model complexities to describe the analyzed beam. The parameter space dimension for higher-order mode systems rapidly increases when considering models using an increasing mode count. A typical solution is to limit the number of modes by defining a limit or manually selecting the modes expected to be output by the laser system. 
    
    This paper presents an algorithm that fits a user-defined set of Hermite- or Laguerre- Gaussian modes to an experimental beam by determining the beam parameter and mode's complex amplitudes. Its required input is a series of intensity pictures by a CCD or CMOS camera taken over several distances along the beam's propagation axis, ideally more than one Rayleigh range. The software uses a user-defined set of Hermite- or Laguerre-Gaussian modes as a basis of its theoretical model and a non-linear minimizer to search the mode's parameter space. A novel technique is presented to significantly improve the computation times of the Hermite-Gaussian complex amplitudes, which is used in combination with other known improvements. This way, even mode bases containing more than sixty Hermite- or Laguerre-Gaussian modes can be computed on standard PCs within hours (as of 2023).
    
    First, we briefly introduce the algorithm and then discuss its mathematical foundation in \Cref{sec_algorithmtheory}. Next, the computational improvements are described in \Cref{algopt}. A full description of the algorithm's processing flow and some first experimental comparisons can also be found in the Appendix (\Cref{sec_appendix}). The software will be open-sourced soon, so the source code and program will be freely available under a public license. Until then, the source code will be available upon request.
    
    \section{The fit algorithm}
    \label{thealgorithm}
	This algorithm aims to describe a real laser by a mathematical model by decomposing the laser beam intensity profile into a linear combination of Hermite- or Laguerre-Gaussian modes. A non-linear minimizer searches said beam model's parameter space to match the experimental data as closely as possible to the measured intensity profiles. For all intents and purposes, the minimizer in this paper will be considered a black box that can sufficiently minimize a value by exploring a multi-dimensional parameter space. The one used here is from our in-house interferometer simulation tool, IfoCAD, but in theory, any other minimizer could be used here instead.
 
    The algorithm fits the input data into a theoretical model. Therefore, we will look at the fundamental idea of describing laser beams by their intensity profiles using laser modes. Then, we will look at the optimization steps to drastically reduce the computational cost to calculate the beams model and describe the sequential order in which the algorithm fits data.
    
    \subsection{Creating beam models from intensity profiles}\label{sec_algorithmtheory}
        This section will look at the fundamental idea of using intensity profiles to derive mathematical models using fitting algorithms. This idea is familiar and was already successfully performed in \cite{Ma2021, Kim2021, Choi2019, Huang2015}.

        An electric field distribution can be described as a superposition of modes as
	\begin{equation}
    \label{eq_model}
		  E_\text{model}(x,y,z) = \sum_{m=0}^{k}\sum_{n=0}^{l}\alpha_{mn}E_{mn}(x,y,z),
	\end{equation}
 
        with $\alpha_{mn}$ being the complex amplitudes and $E_{mn}$ the electrical field spacial distribution for a mode with indices $m,n$. The wavelength $\lambda$ is assumed to be known and fixed. All modes are defined based on a common fundamental mode $E_{00}$ with fixed beam axis along $z$, waist position $z_0$, and Rayleigh range $z_R$. The latter two are essential fit parameters and will be determined by this method. It is searched for a single beam model that describes all intensity profiles simultaneously. For the linear combination of such fields $E_{mn}$, the algorithm uses linear combinations of either the Hermite- or Laguerre-Gaussian mode basis.	These sets of modes are complete sets of orthogonal functions. Any linear combination containing such modes is a solution to the Helmholtz equation. They are defined as

    \begin{align}
        H_{mn}(x,y,z) =& \frac{1}{\sqrt{2^{m+n-1}\,\pi\, m!\,n!}}\frac{1}{w(z)} 
		H_m\left(\frac{x\sqrt{2}}{w(z)}\right) H_n\left(\frac{y\sqrt{2}}{w(z)}\right)
		\nonumber\\ &\cdot 
    \label{eq_hg}
		\exp\left(-\frac{r^2}{w(z)^2}\right)
		\exp\left(\text{i}\frac{kr^2}{2R}\right)
		\exp\left(\text{i}(m+n+1)\eta(z)\right),
    \end{align}
	and
	\begin{align}
		{LS}_{mn}(r,\varphi,z)=&2\,\sqrt{\frac{m!}{(1+\delta_{0,n})\pi(m+n)!}}
		\frac{1}{w(z)} \sin (n\cdot\varphi)
		\left(\frac{r\sqrt{2}}{w(z)}\right)^n 
		L_m^n\left(\frac{2r^2}{w(z)^2}\right)
		\nonumber\\&\cdot 
		\exp\left(-\frac{r^2}{w(z)^2}\right)
		\exp\left(\text{i}\frac{kr^2}{2R}\right)
		\exp\left(\text{i}(2m+n+1)\eta(z)\right),
		\label{eq_ls}\\
		{LC}_{mn}(r,\varphi,z)=&2\,\sqrt{\frac{m!}{(1+\delta_{0,n})\pi(m+n)!}}
		\frac{1}{w(z)} 
		\cos (n\cdot\varphi)
		\left(\frac{r\sqrt{2}}{w(z)}\right)^n
		L_m^n\left(\frac{2r^2}{w(z)^2}\right)
		\nonumber\\\label{eq_lc}&\cdot 
		\exp\left(-\frac{r^2}{w(z)^2}\right)
		\exp\left(\text{i}\frac{kr^2}{2R}\right)
		\exp\left(\text{i}(2m+n+1)\eta(z)\right)
		,
	\end{align}
        where $w(z)$ is the beam radius at position $z$, $r = \sqrt{x^2+y^2}$ is the radial distance from the beam's center, $\eta(z)$ its Gouy phase, and $R$ the radius of curvature of the wavefront. The Hermite-Gaussian mode description \Cref{eq_hg} uses Hermite polynomials $H_n$ to describe the electric fields. In the Laguerre-Gauss modes (\Cref{eq_ls} and \Cref{eq_lc}), the generalized Laguerre polynomials $L^n_m$ are used, with the azimuthal index $n$ and radial index $m$. The set $\{H_{mn}\}$ is orthogonal and complete, as is the set $\{LS_{mn}, LC_{mn}\}$. While these two sets are independent, their modes still can be represented as linear combinations of modes from the other set \cite{Kimel1993, ONeil2000}.
 
    Using linear combinations of modes from one of the two mentioned sets can resemble any given individual intensity profile. Nevertheless, these solutions are not unique, and more than one mode and amplitude combination can be found to resemble one single intensity distribution. However, the different solutions possess different propagation behavior. The ambiguity is resolved by fitting intensity profiles from different propagation distances to one single beam model. The algorithm must, therefore, be fed with at least three pictures along the beam's propagation axis over a distance of usually at least one Rayleigh range. All of these beam profiles are fitted to one single common beam model. This way, the model's uniqueness can be assured \cite{Xue2000}. 
    
    Besides the beam parameters that need to be determined by the fit, a set of setup-related parameters must be considered for every image. It is impractical to take pictures precisely along the propagation direction while maintaining the exact positioning of the camera relative to the beam axis, and the true beam center position on the sensor would still be unknown. Therefore, parameters for the beam's center position on the sensor are needed in each picture's $x$ and $y$ directions and are referred to as $x_0$ and $y_0$. These parameters are defined as the offset from the bottom left corner in each picture and range from zero to the maximum pixel count of the camera in the x and y direction with sub-pixel accuracy. Also, a non-zero background illumination of the sensor is considered. For this, a parameter $b$ is introduced as a constant or linear offset for the model. Lastly, not each picture has the same gain or exposure time while recording the intensity profile. That gives a varying scaling between pictures and is modeled by the scaling factor $s$, a real-valued constant number. These additional parameters leave us with a parameter space that rapidly increases with the number of pictures we provide. Our algorithm resolves this problem by numerically calculating the setup parameters, $x_0$, $y_0$, $b$, and $s$, explicitly outside the core non-linear minimization step, described in \Cref{algopt}. The resulting model is
	\begin{align} \label{eq_intensitymodel}
		y^\text{model}_{ij} = s_i\cdot|E_\text{model}(x_j-x_{0_i},y_j-y_{0_i},z)|^2+b_i,
	\end{align}
    and describes the intensity at each pixel position $j$ on the image number $i$. The setup parameters are the scaling factor $s_i$, the background offset $b_i$, and the beam center position $x_{0_i}$ and $y_{0_i}$. $E_\text{model}$ is a superposition of either Hermite- or Laguerre-Gaussian modes as defined in \Cref{eq_model}, meaning the $E_{mn}$ is either chosen from the set $\{H_{mn}\}$ or $\{LS_{mn}, LC_{mn}\}$. The free parameters are implicitly contained in there, which are the Rayleigh range $z_R$, waist position $z_0$, and a complex amplitude to each higher order mode $\alpha_{mn}$. To resolve another ambiguity, the coefficient $\alpha_{00}$ of the fundamental mode is fixed to $\alpha_{00} = 1 + i0$. By doing so, each other amplitude value is expressed in relative terms to the 00-mode, having the added benefit of removing this amplitude as a free parameter from the fitting process and reducing the number of free parameters by 2. 
    
    The total number of parameters for the model is
	\begin{equation}
		n_\text{parameters} = 
		\overbrace{\underbrace{2}_{z_0,z_R}+
		\underbrace{2\cdot(n_{\rm modes}-1)}_{\rm Re, Im}}^{\text{true parameters } p}+
		\overbrace{\underbrace{4\cdot n_{\rm image}}_{x_0,y_0,s,b}.}^{\text{setup parameters } q},
	\end{equation}
    and they can be divided into two distinct parameter sets: $p$, which we will refer to as true parameters because their parameter space is searched by a non-linear minimizer, and setup parameters $q$, which are solved independently and therefore not handed to the minimizer. Also, the number of parameters for the amplitudes is reduced by one due to fixing the complex amplitude of $\alpha_{00}$.
    
    Some important degeneracies between fit parameters and higher-order modes must be considered here. The modes $H_{0,1}, H_{1,0}$ are degenerate in first order with the $x_0$ and $y_0$ parameters. Also, the LG$_{10}$, a doughnut-shaped mode, which is a linear combination of the modes $H_{0,2}$, $H_{1,1}$, and $H_{2,0}$ is degenerate with the Rayleigh range $z_R$ and waist position $w_0$ \cite{Bond2016}. Thus, we cannot use the LG$_{10}$ mode in our beam models and can only estimate the $H_{0,1}, H_{1,0}$ mode if $x_0$, $y_0$ are well constrained.
   
    To compare experimental data with the fitted beam model, our figure of merit, or error function, is the Mean Square Error (MSE) difference between the images from the CCD cameras and its internal beam model,
	\begin{equation}\label{eq_errorfunc}
		\text{MSE} = 10^6\frac{1}{n_\text{image}}\sum_{i_\text{image}=0}^{n_\text{image}}\left(\frac{1}{n_\text{pixel}}\sum_{j\in \text{pixel}}^{n_\text{pixel}}(y_{ij}^\text{data}-y_{ij}^\text{model})^2 \right)
	\end{equation}
	where $y^\text{data}_{ij}$ is the intensity at a pixel of index $j$ of the CCD image $i$, the sum gets normalized by the number of pixels $n_\text{pixel}$ of the sensor and number of images $n_\text{image}$ used. A scaling factor of $10^6$ brings the resulting MSE into a convenient order of magnitude around 10. The $y_{ij}^\text{model}$ are the values at the point $ij$ given by the mathematical model described in \Cref{eq_intensitymodel}. This cost function is minimized by the algorithm's non-linear minimization step by searching the parameter space of the complex amplitudes $\alpha_{mn}$ together with the beam parameters $z_R$ and waist $w_0$. This way, the difference between the chosen mode combination and the experimental beam profiles provided will be reduced as far as possible by the chosen non-linear optimizer.
 
    The next part of this section will describe steps that significantly improve the computational time needed, which is prohibitive if the equations above are directly applied to a nontrivial set of pictures and modes.
    
	\subsection{Runtime optimization} \label{algopt}
\begin{figure}[ht!]
		\centering
		\includegraphics[width=0.80\linewidth]{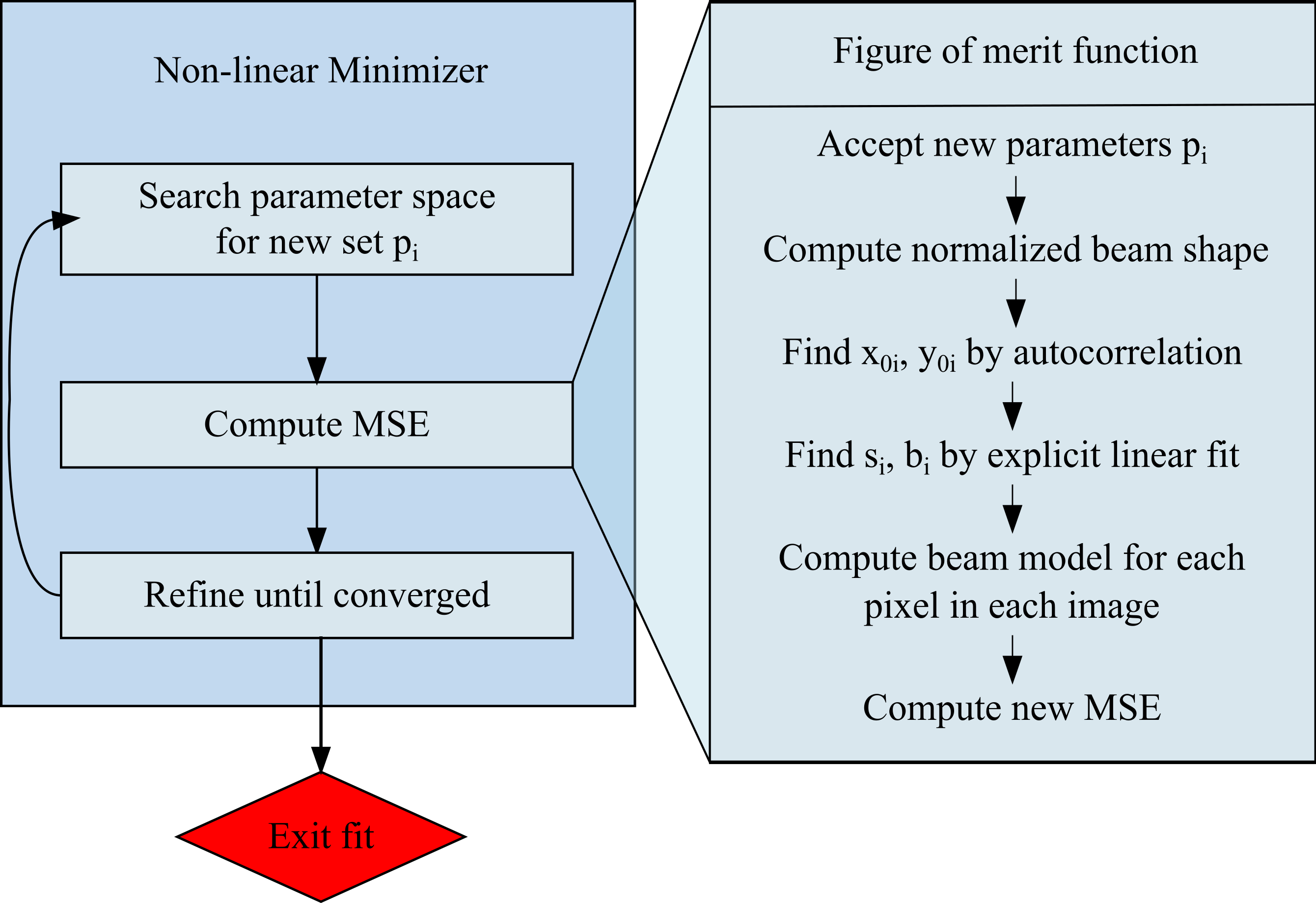}
		\caption{A visualization of the steps within the minimization routine. It repeats the three main steps: Searching the parameter space using the non-linear minimizer using the Mean Square Error (MSE) as a figure of merit. The new parameter set $p_i$ then is used in the calculation for a new MSE value. Lastly, it is checked if the algorithm converged, and these steps are repeated otherwise. The calculation of the MSE value has some additional steps to reduce its runtime. First, an orthogonal decomposition of the normalized beam shape reduces the number of calculations done. Then, the parameters for the beam center $x_0$ and $y_0$, scaling factor $s$, and background $b$ are pre-calculated for each intensity profile, so they are not part of the non-linear minimization. Lastly, the beam model is calculated for each pixel, where the number of calculations can be reduced with a technique similar to that used for the beam shape.}
		\label{beamfitmsesteps}
	\end{figure}

        This section will describe the routine around the nonlinear minimization in greater detail and measures taken to reduce the computational effort needed. That allows the algorithm to run on regular machines within minutes to hours. The routine is illustrated in \Cref{beamfitmsesteps}.
	
        The algorithm is optimized in two main ways: First, by optimizing the time it takes to compute the model's new intensity profiles, and second, by reducing the number of parameters handled by the nonlinear minimizer. In addition, the workload is distributed by parallelization, where one thread is created per image.
        
        When the non-linear minimizer creates a new set of parameters $p_i$, a few steps must occur to arrive at the new figure of merit. The algorithm needs to re-calculate the intensity for each pixel on each picture using the new waist radius and position. These calculations can be highly sped up using Hermite-Gaussian modes rather than Laguerre-Gaussian modes. Camera pixels are arranged on a rectangular grid, which matches the symmetry of the Hermite Gaussian modes. This can be exploited to reduce the computational effort, as described in the following paragraphs.
        
        All Hermite-Gauss modes $H_{m,n}$ are conventionally calculated individually and then weighted by their complex coefficients $\alpha_{mn}$ to calculate the intensity for a given mode combination in each pixel at position $(x,y)$ for a picture at position $z$, according to \Cref{eq_model} and \Cref{eq_intensitymodel}.	To fit a model using the modes from order $(0,0)$ up to a combined order of $(m,n)$ with $m + n \leq N$, the number of Hermite-Gaussian modes to calculate is $\frac{1}{2}N(N+1)$ and therefore the number of these calculations scales with $\mathcal{O}(N^2)$. To arrive at a problem with less complexity, first \Cref{eq_hg} has to be split into factors depending on $m$ or $n$, respectively.
	\begin{subequations} \label{eq_hgsplit}
		\begin{align}
			\label{cfactor}
			H_{mn}(x,y,z) =& \sqrt{\frac{2}{\pi}}\frac{1}{w(z)} \exp\left(-\frac{r^2}{w(z)^2}\right) \exp\left(i\frac{kr^2}{2R}\right)\\
			\label{nfactor} &\cdot\frac{1}{\sqrt{2^mm!}}H_m\left(\frac{x\sqrt{2}}{w(z)}\right) \exp\left(-i(m+\frac{1}{2})\eta(z)\right)\\
			\label{mfactor} &\cdot\frac{1}{\sqrt{2^nn!}}H_n\left(\frac{y\sqrt{2}}{w(z)}\right) \exp\left(-i(n+\frac{1}{2})\eta(z)\right)
		\end{align}
	\end{subequations}
        This way, the factors (\ref{nfactor}) and (\ref{mfactor}) for a given $n$ or $m$ are calculated individually and only once for each respective index $m$ and $n$. Also, the factor \ref{cfactor} is computed only once for all modes in their respective pixel position. The results can be used to compute the other modes depending on the same index without an additional calculation. This way, the lengthy polynomial calculations are only done once for each index, and individual modes only require simple multiplications. With this, only a total of $2N$ factors have to be computed instead of $\frac{1}{2}N(N+1)$.
 
        The calculation time can be further reduced by applying the same approach to compute the intensities for each pixel position on the screen. For this, \Cref{eq_hgsplit} is further divided into terms of $x$ and $y$ coordinates.
	\begin{subequations} \label{eq_hgsplitxy}
		\begin{align}
			H_{mn}(x,y,z) =& \sqrt{\frac{2}{\pi}}\frac{1}{w(z)}\\
			\label{x_factor}
			&\cdot\frac{1}{\sqrt{2^mm!}}H_m\left(\frac{x\sqrt{2}}{w(z)}\right) 
			\exp\left(-i(m+\frac{1}{2})\eta(z)\right)
			\exp\left(-\frac{x^2}{w(z)^2}\right) \exp\left(i\frac{kx^2}{2R}\right)\\
			\label{y_factor} &\cdot\frac{1}{\sqrt{2^nn!}}H_n\left(\frac{y\sqrt{2}}{w(z)}\right)
			\exp\left(-i(n+\frac{1}{2})\eta(z)\right)
			\exp\left(-\frac{y^2}{w(z)^2}\right) \exp\left(i\frac{ky^2}{2R}\right)
		\end{align}
	\end{subequations}
        Now, the factors (\ref{x_factor}) and (\ref{y_factor}) have to be computed only once for each possible pixel position in $x$ or $y$ direction, respectively. That also reduces the number of complex polynomial calculations from an amount equal to the number of pixels squared to only two times the number of pixels. Simple multiplications can again replace the other calculations.
        
        Using Laguerre-Gauss modes for fits is still possible because they can be expressed as a combination of Hermite-Gauss modes via basis transformation. The relations between these modes are mathematically exact and are described in greater detail in a paper from Kimel \cite{Kimel1993} or O'Neil \cite{ONeil2000}. Each Laguerre-Gauss mode can be expressed as a linear combination of Hermite-Gauss modes, which can then be efficiently calculated as described above. The program uses only Hermite-Gauss modes for its internal calculations, while the user can choose Laguerre-Gaussian modes for their models. The software automatically and transparently handles the transformations between bases.
        
        Next, we reduce the number of free parameters the optimizer has to fit by calculating the model's setup parameters, which are the scaling factor $s_i$, a non-zero background $b_i$, and the beam's center position relative to the center of the screen $x_{0_i}$, $y_{0_i}$. Their nature allows us to solve for them explicitly for each picture individually. They are calculated anew for each new set of true parameters $p_i$ produced by the minimizer. Their calculation also happens in parallel for each image $i$.
        
        As noted in \Cref{eq_model}, the scaling factors $s_i$ and backgrounds $b_i$ occur linearly in our beam model. A simple linear regression finds exact results in one step without iteration, which are then used to calculate the new figure of merit. If needed, the background can even be represented as a linear slope of the form $b_i = (a_1x+a_2y+a_3)_i$ instead of a constant background and still be solved for.
        
        Estimating the beam center $x_{0_i},y_{0_i}$ in each image uses a cross-correlation between the intensity profile calculated with the new set of parameters and the measured intensity profile. This correlation is done after each iteration of the minimizer and for each individual picture. A numerical cross-correlation $c$ can quantify the similarity between each image's measurement and simulation data. It is efficiently computed with a 2-D discrete Fourier transform $\mathcal{F}$ by
	\begin{align}\label{eq_crosscorr}
		c(\tau) = \mathcal{F}^{-1}(\overline{\mathcal{F}(\text{data})}\cdot\mathcal{F}(\text{model})),
	\end{align}
        where $\tau$ is the estimated distance between the model and data, and $\overline{\mathcal{F}}$ is the complex conjugate of $\mathcal{F}$. Matching templates with cross-correlation is well understood and can be found in \cite{Briechle2001} and \cite{Lewis2001}. The fast Fourier transformations are calculated pixel by pixel using the highly optimized FFTW3 software package. This way, $x_0$ and $y_0$ are found as the 2-D displacements where the correlation is maximal. An interpolation is used to achieve a sub-pixel accuracy for the beam center estimate. This interpolation is performed around the discrete maximum of the cross-correlation. A bi-quadratic function is used to model the maximum and points in this local area to reach a sub-pixel accuracy. The resulting function has the form	
	\begin{align}\label{eq_crosscorr_biquadrat}
		c(x,y) = a_0+a_xx+a_yy+a_{xx}x^2+a_{yy}y^2+a_{xy}xy,
	\end{align}
        where the parameters $a_j$, with $j \in (0,x,y,xx,yy,xy)$, are chosen in order to have a smooth function $c(x,y)$ to approximate the maximum. Solving for said functions maximum in $x$ and $y$ direction, we obtain $x_0$ and $y_0$ with sub-pixel resolution. These parameters are then used to calculate the new cost function defined in \Cref{eq_errorfunc} using the beam model, which is efficiently calculated using the Hermite-Gaussian mode factorization as described before.
        The new value for the figure of merit is then returned to the minimizer, which can continue searching the free parameter space and produce a new set of beam parameters $p_i$, which starts the chain of calculations again.

    \begin{figure}[ht!]
    \centering
    \begin{minipage}[t]{0.48\textwidth} 
        \includegraphics[width=0.98\linewidth]{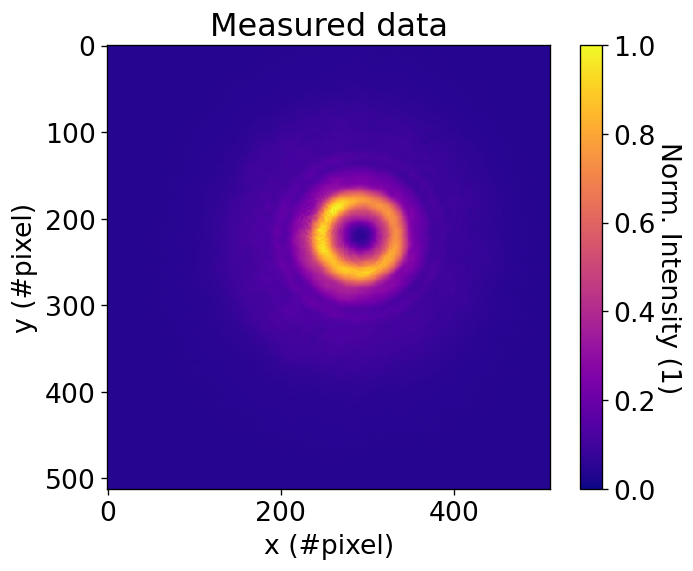}
        \caption{Example CCD camera image of an experimental beam with significant deviation from a fundamental Gaussian beam. The x- and y-coordinates are given in pixel count, with a pixel pitch of \qty{5.5}{\micro\meter}. The intensity is normalized to the highest value in the picture.}
        \label{fig_experimentalbeam}
    \end{minipage}
    \end{figure}

    \begin{figure}[ht!]
    \begin{minipage}[t]{0.48\textwidth}
        \centering
        \includegraphics[width=0.98\linewidth]{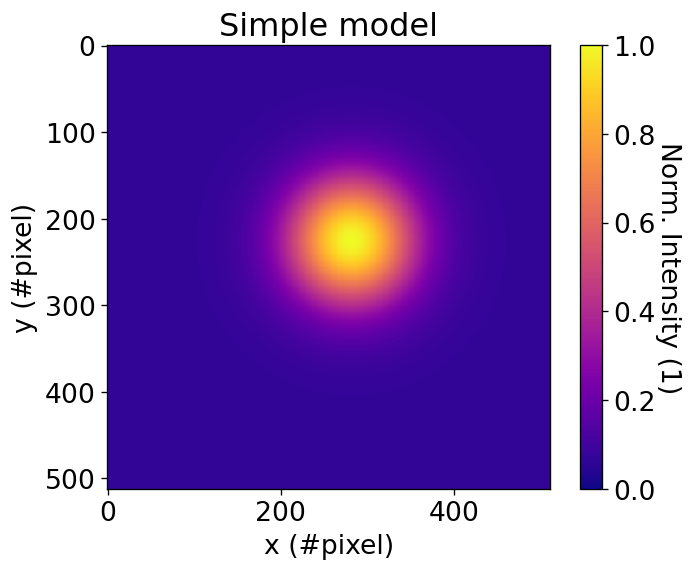}   
    \end{minipage}
    \begin{minipage}[t]{0.48\textwidth}
        \centering
        \includegraphics[width=\linewidth]{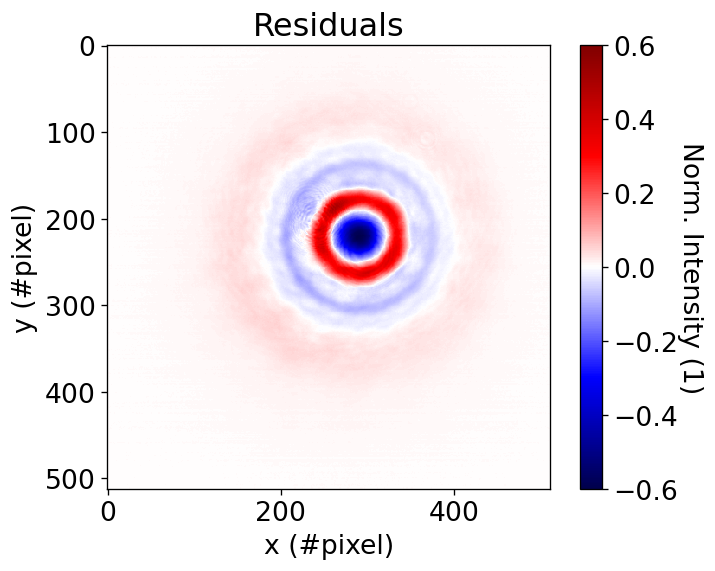}
    \end{minipage}
    \caption{A linear data fit using only the fundamental Gaussian mode. The simple models intensity picture (left) is normalized to the same scale as \Cref{fig_experimentalbeam}. The residual picture (right) is a pixel-by-pixel difference between the experimental data in \Cref{fig_experimentalbeam} and the depicted intensity profile on the right. The red regions are where the intensities of the experimental data are higher than the model's prediction, and for the blue areas, it is vice versa. Maximum discrepancies reach nearly values of $\pm0.6$, equating to a maximum difference of \qty{60}{\percent}.}
    \label{fig_fundamentalfit}
    \end{figure}
    
    \begin{figure}[ht!]
        \begin{minipage}[t]{0.48\textwidth}
        \centering
        \includegraphics[width=0.98\linewidth]{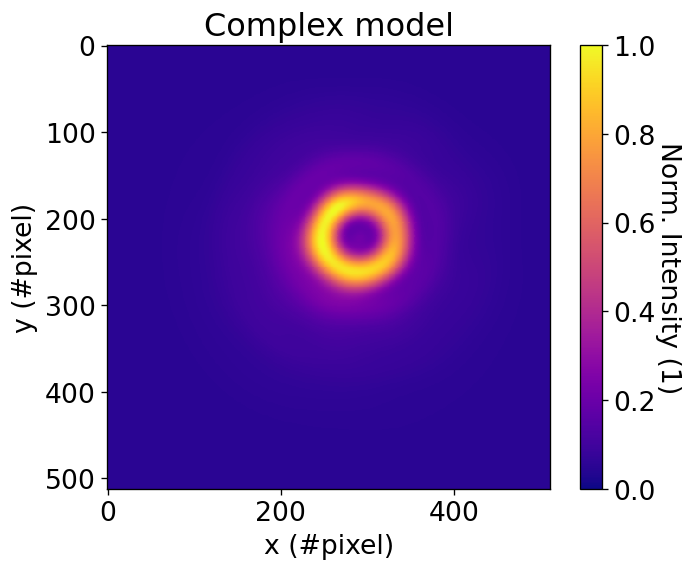}
    \end{minipage}
    \begin{minipage}[t]{0.48\textwidth}
        \centering
        \includegraphics[width=\linewidth]{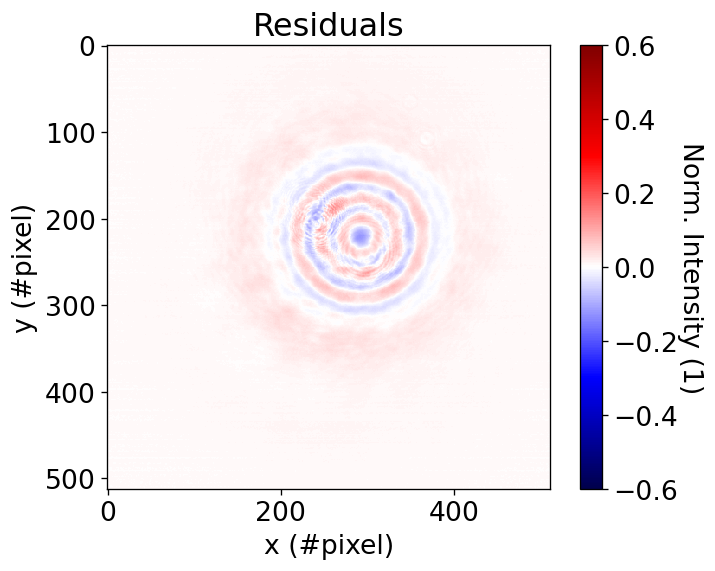}
    \end{minipage}
    \caption{A fit using a mode basis containing 76 Hermite-Gaussian modes. The same scale is used here as in \Cref{fig_fundamentalfit}. In the residual picture on the right, the maxima in the image reach values of $\pm0.1$, which equates to a maximal discrepancy of \qty{10}{\percent} between the beam model and measurement data. Ring-shaped structures are clearly visible in the residual images.}
    \label{fig_higherorderfit}
    \end{figure}
        
        In addition to the core minimizing function described above, the program contains a few auxiliary steps necessary for practical use. These include cropping and downsampling the input images to reduce their number of pixels, identification of bad pixels, and graphical output of the model's fit and residuals as data files and PNG image files. Some additional information can be found in the appendix (\Cref{sequential}). A more comprehensive and complete description can be found in the Beamfit software user manual, which will be available with the release of this software.

        Using the above-described algorithm, the Beamfit software produces a model that resembles the measurement data as closely as possible. An example output is shown through \Cref{fig_experimentalbeam,fig_fundamentalfit,fig_higherorderfit}. In \Cref{fig_experimentalbeam}, one example of an experimentally recorded intensity profile is shown, which is part of a series of images used for the Beamfit algorithm. Then, \Cref{fig_fundamentalfit} shows a linear fit only using the fundamental Gaussian mode. Evident discrepancies between the model and data can be spotted between the intensity profiles and are even more prevalent in the residual picture. In \Cref{fig_higherorderfit}, a model of 76 Hermite-Gaussian modes was used to fit the experimental data set. The resulting intensity distribution resembles the experimental data in \Cref{fig_experimentalbeam} much closer. However, clear structures are still visible in the residuals, hinting that modes are still missing from the model.

        In addition to the intensity profiles and residual pictures, the algorithms also put out the normalized complex amplitudes used to produce the beam model. They can then be used in further simulations or other post-processing steps. The Beamfit software succeeds in its task of resembling the measured data as closely as possible. This model describes the provided data at hand; however, it is important to note that this does not necessarily predict the correct propagation behavior of the fitted beam. To show that, additional experimental verification is needed.

    \section{Summary}
    We introduced the Beamfit algorithm that uses a non-linear minimizer to create a theoretical model of experimental laser beams. In \Cref{thealgorithm}, we presented the algorithm working principle and ways to reduce the computational time needed significantly. First, we showed that the number of Hermit-Gaussian mode calculations performed for the complex amplitudes and intensity values at each pixel can be significantly reduced from orders of $\mathcal{O}(N^2)$ to $\mathcal{O}(N)$. It was achieved by separating the mode into two distinct factors, as described in \Cref{eq_hgsplit}, where they depend only on $m$ or $n$, respectively. Then, the same principle is used to calculate the intensity values for each pixel. Here, the mode gets factored into their $x$ and $y$ coordinates, respectively, as described in \Cref{eq_hgsplitxy}. The parameter space handed to the minimization routine was also reduced by pre-calculating the parameters needed to fit the mathematical model to experimental data. The scaling factor $s$ and background $b$ can be solved linearly, and the on-screen position of the beam's center was calculated by a cross-correlation between the mode-model and CCD camera data, achieving sub-pixel accuracy by using interpolation.
    
    \section{Appendix}\label{sec_appendix}
    
    \subsection{Beamfit pre-and post processing} \label{sequential}
    When invoking the algorithm, the user can provide four inputs:
    \begin{itemize}
        \item Setup file
        \item CCD camera pictures in text format
        \item Include text file containing precomputed parameters
        \item List of bad pixels
    \end{itemize}
    The first two on the list are required to run the algorithm, and the last two are optional inputs.
    As a setup file, the algorithm requires an ASCII text file containing lines of instructions needed to run it. The user provides the program with crucial information using this file, like the system of modes to use for fitting. Additional optional instructions are also provided here.

    The second required input is a series of intensity-profile pictures. An ideal set contains ten or more images and spans at least one Rayleigh range of the given beam with a fixed reference point along the beam path.

\begin{figure}[ht!]
		\centering
		\includegraphics[width=0.97\linewidth]{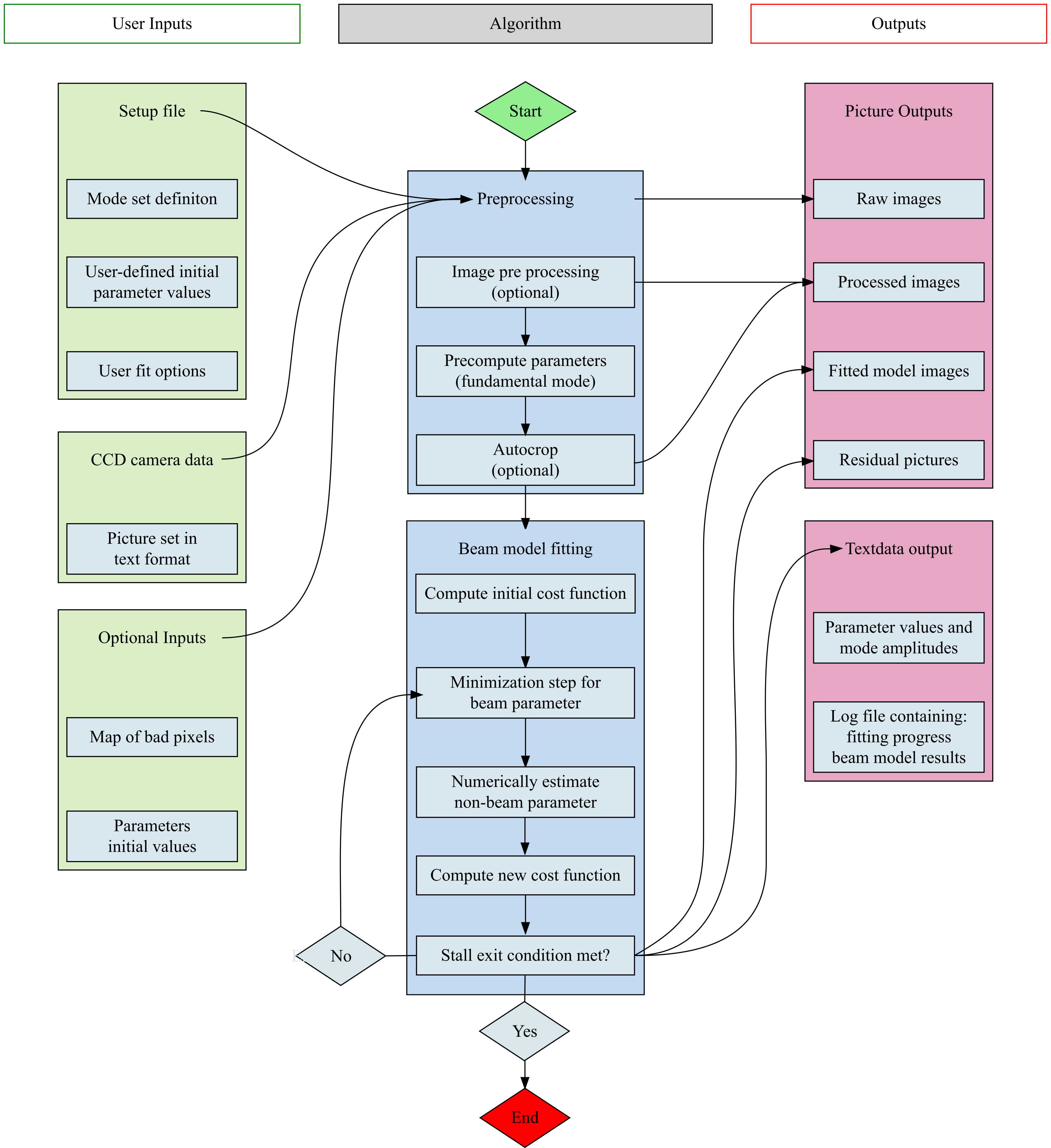}
		\caption{This flowchart illustrates the beam fitting algorithm's order of operation with all its in- and outputs. On the left side, all possible inputs to the algorithm are listed. The first two blocks are required to run the algorithm. The middle part shows the steps run through by the algorithm. Depicted are the start and stopping points and two main blocks. The first contains pre-processing steps to prepare the data for the next block. In the second, the main part of the algorithm gets invoked to fit a user-chosen beam model to the input picture set. Finally, on the right side, the different outputs are shown produced by the algorithm. Data gets put out in some steps of the process. Log and include files get created at the end of each model fitting step.}
		\label{beamfitflowchart}
	\end{figure}

    Some pre-processing steps prepare the CCD data before the main model fitting can occur, as seen in \Cref{beamfitflowchart}. These steps are often essential to further reduce the calculation time of this algorithm significantly by down-sampling and cropping the images, especially when the model of choice contains many modes.

    \textbf{Image pre-processing:} Multiple pre-processing routines are part of the software, which users can invoke as needed. They are invoked in the order of cropping, handling bad pixels, and then down-sampling.

    \textbf{Precompute parameters:} Initial guesses for the beam parameters using only the TEM$_{00}$ mode. Estimated are the Rayleigh range $z_r$ and the waist position $z_0$.

    \textbf{Autocrop:} The pre-fit can additionally serve in an auto-cropping routine to estimate the size of the beam, which cuts out a rectangular area around the estimated beam center. The rectangle's size is scaled with the beam's estimated diameter.

    After the initial image pre-processing step, the algorithm starts with its primary function: fitting the user-chosen beam model to the provided beam data using the non-linear minimization routine.

    \textbf{Compute initial cost function:} The first step is the initial calculation of the cost function, as defined in \Cref{eq_errorfunc}. The model uses the beam parameter values of the fundamental mode calculated during the pre-calculation step. This first step gets invoked only during the first execution of this block.

    \textbf{Minimization step for beam parameters:} Next, the minimization step gets invoked. It uses a non-linear optimizer that is a combination of a particle swarm optimizer, a modified Nelder-Mead simplex, and a modified Levenberg-Marquad algorithm, which are inherited from LISO and IfoCAD. It provides the set of parameters from the pre-fit or the previous "compute new cost function" step. Then, it calculates a new set of beam-model parameters, which are handed down to the next step.

    \textbf{Numerically estimate non-beam parameter:} Here, the algorithm calculates the setup-parameters $s$, $b$ ,and $x_0$ and $y_0$ as described in the last part \Cref{algopt}.

    \textbf{Compute new cost function:} At this point, a complete set of new parameters for the model has been calculated and can be used to calculate the new value for the figure of merit.

    \textbf{Exit condition met:} Here, the minimizer's exit condition is checked. When it is met, the whole algorithm concludes. If not, the algorithm returns to the minimization step with the new set of parameters and the updated figure of merit to run again.
    All calculated parameters get exported in different formats during this step, whether the exit condition gets met. The program exports two picture sets, one with the calculated intensities and the other containing a pixel-by-pixel difference between the measured data and the estimated model. Additionally, text files containing the algorithm's progress and the estimated parameter values are created.
    
	\section*{Acknowledgements}
	We thank Nina Bode for providing access to the diagnostic breadboard and her extensive advice and support on its operation. Also, we would like to thank Tim Haase for his support in delivering a better software infrastructure and code maintenance for the beam-fitting algorithm. Another thanks goes to Matthew Scourfield for his advice on confidence for model fitting and data analysis.\\
	
	\noindent The authors would like to acknowledge the Deutsche Forschungsgemeinschaft (DFG, German Research Foundation) for funding the SFB TerraQ (Project-ID 434617780 – SFB 1464) and QuantumFrontiers (EXC 2123, Project ID 390837967), as well as the Max Planck Society (MPG) in the framework of the LEGACY cooperation on low-frequency gravitational wave astronomy (M.IF.A.QOP18098, CAS’s Strategic Pioneer Program on Space Science XDA1502110201).
	
	\noindent JJC and SMK would like to acknowledge funding in the framework of the Max-Planck-Fraunhofer Kooperation Project Glass Technologies for Einstein Telescope (GT4ET).

    \section*{Funding}
    This work was funded by the Deutsche Forschungsgemeinschaft (DFG, German Research Foundation) in the context of the SFB TerraQ (Project-ID 434617780 – SFB 1464) and QuantumFrontiers (EXC 2123, Project ID 390837967), by the Max Planck Society (MPG) in the framework of the LEGACY cooperation and the Max-Planck-Fraunhofer Kooperation Project Glass Technologies for Einstein Telescope (GT4ET).
	
	\section*{Conflict of interests}
	The authors declare no conflicts of interest.

\bibliography{beamfit}

\end{document}